# Exploring the relation between students' online learning behavior and course performance by including contextual information in data analysis

Zhongzhou Chen*, Mengyu Xu, Geoffrey Garrido, and Matthew W. Guthrie

*Department of Physics, University of Central Florida, 4111 Libra Drive, Orlando, Florida, 32816*
*Department of Statistics and Data Science, University of Central Florida, 4000 Central Florida Blvd., Orlando, Florida, 32816*

This study examines whether including more contextual information in data analysis could improve our ability to identify the relation between students' online learning behavior and overall performance in an introductory physics course. We created four linear regression models correlating students' pass-fail events in a sequence of online learning modules with their normalized total course score. Each model takes into account an additional level of contextual information than the previous one, such as student learning strategy and duration of assessment attempts. Each of the latter three models is also accompanied by a visual representation of students' interaction states on each learning module. We found that the best performing model is the one that includes the most contextual information, including instruction condition, internal condition, and learning strategy. The model shows that while most students failed on the most challenging learning module, those with normal learning behavior are more likely to obtain higher total course scores, whereas students who resorted to guessing on the assessments of subsequent modules tended to receive lower total scores. Our results suggest that considering more contextual information related to each event can be an effective method to improve the quality of learning analytics, leading to more accurate and actionable recommendations for instructors.

## I. Introduction

Online learning platforms provide a rich variety of data on students' learning behavior, enabling researchers to explore the relation between learning behavior and learning outcome, motivation, course completion, and other student characteristics. For example, Kortemeyer [1, 2] examined both the relation between frequency of material access and students' course outcome, and the relation between discussion forum posts and learning outcome; Formanek et. al. [3] studied the relation between number of video views, discussion forum participation, peer grading participation and students' level of motivation and engagement in a massive open online course (MOOC); Lin et. al [4] correlated students' access of instructional videos with course performance. In the broader field of learning analytics, more sophisticated analytic methods and algorithms have been developed to either identify patterns in students' online learning behavior [5–8] or predict academic achievement based on large data sets [9–13].

In most of those studies, students' online learning behavior is characterized by the count, frequency or total duration of one or more types of online learning events, such as the number of discussion forum posts or frequency of video views. However, the same type of learning event occurring under different contexts could be generated by distinct types of student learning behavior. For example, a failed problem solving attempt followed by one or more video access or page access events suggests that the student is trying to learn how to solve the problem, while a sequence of failed homework attempts without accessing relevant instructional materials could imply that the student is randomly guessing, especially when the duration of the attempts are short. However, both kinds of failed attempts would contribute equally to the count or frequency of problem attempt data. Gašević et. al. [14] suggested three types of contextual conditions that can have significant impact on learning analytics, based on Winne and Hadwin's self-regulated learning model [15]: **Instruction condition:** such as the course mode, course content, choice of technology, and instructional design. **Internal Condition:** such as the level of utilization of learning tools and the learner's level of prior knowledge. **Learning products and strategy:** including learner's strategy for completing learning tasks, and the quality of the learning product such as annotations or discussion posts.

A number of recent studies have emphasized to varying degrees the context in which online learning events took place, in addition to the number or frequency of events. For example, Wilcox and Pollock [16] examined the impact of four types of contextual information associated with students' answering of online conceptual assessments; Seaton et. al. [17] looked at the impact of the time duration of resource access, Alexandron et. al. and Pallazo et. al [17, 18] utilized time duration and IP address to detect possible copying behavior in students' problem-solving events. Seaton et. al. [20] examined the difference in resource usage that took place when students are completing different tasks in a MOOC.

In this study, we ask the question: *Can outcomes of learning analytics be improved by considering more contextual information associated with each learning event, without increasing the complexity of the analytic methods?* We increased the contextual information associated with each event in three steps, and demonstrated that each step led to increasingly informative descriptions of students' online learning behavior, which enabled us to provide increasingly accurate answers to our second research question: *how do students' overall performance in a physics course correlate with their online learning behavior*? In other words, do students that are often referred to as "struggling" in a physics course study online learning resources differently from those who perform well in the course, and if so, what are the most characteristic differences?

To answer this question, we collected students' online learning data from a sequence of 10 online learning modules (OLMs) which were assigned as homework to be completed over two weeks. Each module contains an instructional component and an assessment component with 1-2 problems. Previous studies have shown that the mastery-based learning design of the OLMs can not only improve student learning outcome [20, 21] but also increase the interpretability and information richness of learning data [23].

The main events analyzed in the current study are the outcomes of each module, as measured by passing, failing, or aborting the assessment component, resulting in 10 events per student. For each pass-fail event, we extracted three types of contextual information: where, when, and how? More specifically:

1. **Where was it:** on which of the 10 modules did each pass-fail event take place?
2. **When did it happen:** did the pass-fail event take place before or after the student accessed the instructional material in each module, and after how many attempts did the student choose to access the instructional material?
3. **How did it happen:** For each pass-fail event, how much time was spent on solving the problems? Multiple previous studies have linked abnormally short problem-solving duration with either random guessing or answer copying [18, 23–27].

Each context corresponds to one of the conditions proposed by Gašević [14]: the "where" reflects the instructional condition of online materials being organized in a sequence of OLMs, the "when" reflects students' internal state of choosing whether to access the learning resources, and the "how" serves as one indication of the strategy of producing the learning product.

We refer to the combination of a pass-fail event and its associated contextual information as an "interaction state," or "state" for short. We created three different levels of interaction states with each level including additional contextual information than the previous level, as explained in detail in section III.B. Therefore, each level contains more states than the previous one.

Students' overall performance in the course is measured by their normalized final course score, which includes scores from homework, two midterm and one final exams, lab activities, and classroom clicker questions. The final course scores directly determine students' letter grade for the course.

Three linear regression models were constructed to associate each of the three levels of interaction states with students' final course score, as well as a baseline model for comparison. To address the issue of collinearity [29] between the large number of variables, we selected for each model a subset of significant variables using a regularized linear regression algorithm LASSO [30], and reconstructed the linear models using those LASSO-selected subsets. Complementary to the linear models, we also plotted students' transition between different states on neighboring modules using a series of parallel coordinate graphs, which is an updated version of the data visualization scheme developed in an earlier study [31]. As detailed in section IV, a complete description of student learning was obtained by combining the linear model with the corresponding parallel coordinate graph for each level.

In section V, we interpret and compare the outcomes of analysis based on the three levels of interaction states, and discuss the benefit of including increasing amounts of contextual information on each event. We show that, in this case, the inclusion of more contextual information results in more informative descriptions of students' learning behavior. The model that includes all three types of contextual information reveals a characteristic difference in the way top and bottom students complete certain OLMs, which provides the most accurate actionable recommendations for instructors. We also discuss the implications of the current results for both instructors and education researchers, as well as caveats and future directions of the current study.

## II. Study Setup

### A. Design of OLM sequence

The OLM sequence is created using the Obojobo learning objects platform, developed as free and open source software by the Center for Distributed Learning at University of Central Florida [32]. Each OLM consists of an instructional component (IC) and an assessment component (AC) (c.f. Figure 1). The AC contains 1-2 multiple choice problems and allows a total of 5 attempts. Each of the first 4 attempts are sets of isomorphic problems assessing the same physics knowledge but with different surface features or different numbers. On the 5th attempt, the same problem on the 1st attempt is presented to the students again. On four of the modules used in the current study, a new set of isomorphic problems is presented to students on each of the first 3 attempts, while the same problems on the 1st and 2nd attempt were repeated on the 4th and 5th attempts. Each IC contains a variety of learning resources, including text, figures, videos, and practice problems, focusing on explaining one or two basic concepts or introducing problem solving skills that are assessed by the problems in the AC. Upon opening a new module, a student must make one attempt at the AC before being allowed access to the IC. From a pedagogical perspective, the required first attempt could improve students' learning from the IC [33], via the "preparation for future learning" effect [34]. From a research perspective, the first attempt serves as a de facto pre-test that can measure students' incoming knowledge of the content.

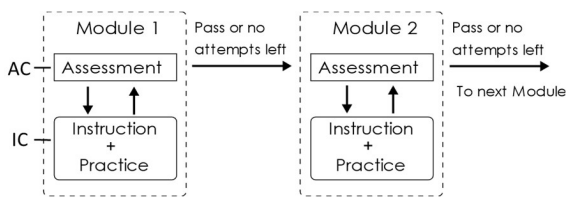

*Figure 1: Schematic illustration of the design of Online Learning Module. Two modules in a sequence are shown in this figure.*

Access to the IC is locked whenever the student is attempting the AC and is unlocked after the answers are submitted. Students are required to access the OLM sequence in the order given. In the 2017 implementation, due to platform limitations, students could access the next module once they had submitted their 1st attempt on the current module. However, students were not explicitly informed of that information, and were encouraged to complete the current module by either passing the AC or using up all attempts before moving on to the next one. We found that in less than 3% of the cases a student accessed the next module in sequence before passing the current one. Those events are excluded from the current analysis.

The OLM sequence used in the current study consists of 10 modules covering the subject of Work and Mechanical Energy. The first six modules introduce the concepts of work, kinetic and potential energy, and conservation of mechanical energy. The AC for those modules consists of mostly conceptual questions (with the exception of module 3, Work and Kinetic Energy). Modules 7-10 focused on solving increasingly complex mechanical energy problems, and the ACs consist of numerical calculation problems. The problems in the AC are inspired by either common homework problems [35] or research-based assessment instruments [36]. Readers can access the OLM sequence following the URL provided in [37]. Due to current platform limitations, all the problems were given in a multiple choice format.

### B. Implementation of OLM sequence

The OLM sequence was implemented in a large calculus-based college introductory physics course in Fall of 2017, taught in a traditional lecture format. Of the 236 students who registered for the class, 184 were male and 52 were female, 107 were ethnic minorities.

The OLM sequence was assigned to students as homework. Modules 1-6 were released one week before modules 7-10, and all 10 modules were due 2.5 weeks after the release of the first six modules. Completing all 10 modules was worth 9% of the total course score, and each module was weighted equally. The modules were released concurrently with classroom lectures on the same topic. The contents of the modules were tested on both the 2$^{nd}$ mid-term exam and the final exam of the course. No other assignments were assigned to the students during the 2.5 week period. A total of 230 students attempted at least one module, and 223 students attempted all 10 modules.

# III. Methods

This section describes in detail how we first extract learning events and contextual information from the raw clickstream data, then integrate the learning events with increasing amounts of contextual information to generate three levels of interaction states for each module, with each level containing one additional type of contextual information. We then describe the three linear regression models created using the three levels of interaction states, plus a baseline model that includes only pass-fail events. We also describe how we address the problem of collinearity within variables using LASSO.

## A. Analysis of students' click-stream data

Students' click-stream data collected from the Obojobo platform are analyzed using R and the tidyverse package [37, 38]. For the current study, we extracted the following types of information from the click-stream data:

*AC attempt outcome and duration:* An AC attempt is recorded as "pass" if the student answers every question correctly, otherwise it is recorded as "fail." The duration of each attempt is recorded as the time between the student clicks a button to start the attempt, and when the student clicks another button to submit the answers.

*Study sessions:* A study session is defined as all students' interaction with the IC between two consecutive AC attempts on a given module. The duration of a single study session is the sum of all the events that took place during the session, including viewing page content and attempting practice problems. Since each module allows a maximum of 5 attempts, and require one attempt before allowing access to the IC, a student can have a maximum of 4 study sessions. However, we observed that in 93% of the cases, each student only had one study session on a given module. In only 6% of the cases did a student have a second study session longer than 60 seconds and at least 30% as long as their longest study session in that module. For those 6% of the cases, we only consider the first of the two study sessions, which is usually the longer one. In the remaining 1% of the cases, the second (and 3rd) study session are neglected because they are either shorter than 30% of the longest study session, or last less than 60 seconds. These choices are unlikely to impact the outcome of the current analysis, because we only consider whether a student had a study session, and how many attempts were made before and after the study session, not the duration of each study session.

## B. Students' Interaction States with OLMs

### 1. Defining interaction states with increasing levels of contextual information

**Level I (3 states):** The first level of interaction states includes information on "**where**" a pass-fail event took place, i.e. whether a student passed or failed the AC of a specific module. We define the following three interaction states for each module:

- **Pass (P):** A student passes the AC within the first 3 attempts. The reason for this choice is that: 1) On four of the modules the AC will provide a different problem only on the first 3 attempts, and will repeat the 1$^{st}$ problem on the 4$^{th}$ attempt 2) Many students do not have the knowledge or skill to pass the AC on their 1$^{st}$ attempt, so they essentially have 2 attempts after studying the IC to be considered as pass, which provides some tolerance for "slips," such as putting in the wrong number in the calculator.

- **Fail (F):** Students who cannot pass the AC within the first 3 attempts. In other words, either passed on the 4th or 5th attempt or failed on all 5 attempts.

- **Abort (A):** Students who did not pass the module and did not use up all 5 attempts before moving on to the next module.

Information of the specific module on which each state occurred is added by combining the module number with the above states when constructing the linear regression model, which is described in detail in section C.1.

Table 1: Definition of Level II states. Note that "Fail" states are defined as failing the first three attemps on each module. The Example column presents the most common event sequence for each state.

| State Name | State Label | Definition | Example |
|---|---|---|---|
| Before Study Pass | BSP | Pass with no study session | {F, P} |
| After Study Pass | ASP | Pass with study session before the 3rd attempt | {F, S, F, P} |
| After Study Fail | ASF | Fail with study session before the 3rd attempt | {F, S, F, F, P} |
| Late Study | LS | Fail with study session after the 3rd attempt | {F, F, F, S, F,P} |
| No Study | NS | Fail with no study session | {F, F, F, F, F} |
| Abort | AB | Fail and did not use all 5 attempts | {F, S, F} |

**Level II (six states):** The second level adds information on "**when**" a passfail event took place with respect to the related study event, on top of the three states in Level I. More specifically, we divided students according to whether their passing or failing of the AC took place before or after studying the IC. Table 1 lists the six states in this level, with examples of common sequences of events belonging to each state, using "S" to represent a study session, and "P" or "F" to represent the outcome of each attempt. All possible event sequences can be categorized into those six states.

The rationale for dividing P and F states according to whether the outcome is achieved before or after the study session is straightforward: students who can pass the module before studying are likely to have stronger incoming knowledge than those who passed after studying. On the other hand, those who studied immediately after the first or second failing attempt likely are more motivated to learn than those who studied after more than 3 failed attempts or did not study at all.

**Level III (nine states):** The third level adds contextual information on "**how**" each pass-fail event is generated by further dividing the BSP, ASP and ASF states according to the duration of the attempts. Different cutoff values have been proposed in several earlier studies to distinguish between an abnormally short attempt and a regular problem solving attempt. In the current analysis, we estimated the cutoff to be 35 seconds, by fitting the attempt duration distribution using scale mixtures of skew-normal distribution models, detailed in the next section. On modules 2 and 6, the cutoffs are adjusted to 17 and 24 seconds respectively on attempts after the study session due to shorter overall attempt durations. We assert that students who spent less than the cutoff times on an AC attempt are unlikely to have put in an authentic effort to solve or even read the problem body.

Table 2: Correspondence between level I, II and III states

| Level I | Level II | Level III |
|---|---|---|
| P | BSP | BSP-N |
| | | BSP-B |
| | ASP | ASP-N |
| | | ASP-B |
| F | ASF | ASF-N |
| | | ASF-B |
| | LS | LS |
| | NS | NS |
| A | AB | AB |

Therefore, we divide each of the BSP, ASP and ASF states into two new states, based on if the students' attempts are classified as "Brief" or "Normal" based on their attempt durations. For example, the BSP state is divided into BSP-B and BSP-N (Before Study Pass-Brief and Before Study Pass-Normal). For BSP and ASP, the attempt duration is taken from the passing attempt which is also the last attempt. On ASF, the duration is taken as the longest of the first 3 attempts. The resulting nine interaction states and the relation between the three levels are listed in Table 2.

**Determining the duration cutoff between Brief and Normal attempts:** Previous studies showed the cutoff between "Brief" and "Normal" attempts can be determined by fitting the distribution of the problem-solving duration using multi-component mixture models (e.g. [40,41]), finding the cutoff between the shortest component and the second shortest component as demonstrated in Figure 2.

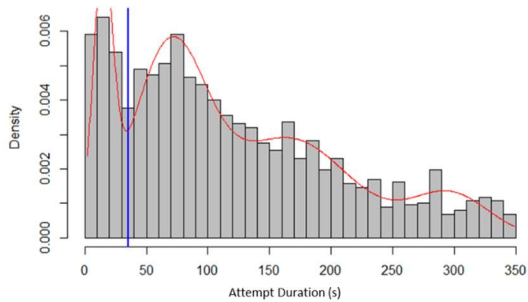

*Figure 2: Example of a mixture model fit of the duration histogram of 1st attempts on all 10 modules combined, with maximum duration of 350 s. The blue line indicating the 35 s cutoff used in the current study.*

In the current study, we fit the distribution of problem solving duration from students' 1st AC attempts collected from all 10 modules, using scale mixtures of normal or skew-normal distributions, since previous studies have suggested that students' problem solving duration distribution are likely skewed [24, 39, 40]. There are two reasons for using the duration data from the 1st attempt. First, because students are required to make their 1st AC attempt before studying the IC, they are more likely to make a random guess. Therefore, the population of brief attempts is more similar to normal attmps, making it easier to separate the short duration component from the rest of the data. Second, on the 1st attempt, students who made a "Normal" attempt must have read the problem text carefully, whereas students who made a "Brief" attempt likely did not, leading to a larger difference in duration between the two components. On their 2nd and 3rd attempts, students may be able to read the problem text faster on some modules where the problems are more similar to the 1st attempt, resulting in smaller difference in duration. For those modules, the cutoff for 2nd and 3rd attempts are being adjusted (see below). The reason for aggregating the duration data from all 10 modules is based on the assumption that "Brief" attempts should be largely independent of the context of the problem, since the student was not actually solving it. Aggregating the duration data will increase the accuracy for estimating the cutoff.

Model fitting is conducted with package mixsmsn [42], and details are presented in the Appendix. Based on the results from model fitting, the cutoff between Brief and Normal attempts is initially set at 35 seconds for all modules. To check if this 35 second uniform cutoff is reasonable for all modules and all attempts, we compared it to the mean of the log duration distribution of attempts made both before and after a study session. We use the mean of log-duration distribution since the distribution is approximately log-normal on many modules. Attempts longer than 7200 seconds are excluded as outliers. For attempts before the study session, the mean log-durations of all modules are between 70 - 200 seconds, much longer than 35 seconds, with harder modules having shorter mean durations. For attempts after the study session, on two modules (2 and 6) the mean log-durations are 35 seconds and 53 seconds respectively; only about half as long as the duration of attempts before study on the same modules. Both modules contain conceptual problems, and the problems presented on the 2nd or 3rd attempt are very similar to the one on the 1st attempt. It is reasonable to assume that students can correctly solve the problem on their 2nd or 3rd attempt by looking at the new diagram and without fully reading the problem body again. Therefore, for those two modules, we treat the shortest 15% of the attempts as "Brief," and adjust the cutoffs to 17 and 24 seconds respectively for attempts after study. On all other modules, the mean duration of attempts after study either increased or decreased slightly (m4). Therefore the same 35 second cutoff is used for all other modules except m2 and m6.

## C. Modeling Linear Relation between Interaction States and Total Course Score

### 1. Initial construction of linear models

We construct three linear regression models between students' interaction states on each module and their final course score for each of the three levels of interaction states, in the form of equation (1), where for the $i^{th}$ student, $y_i$ is the standardized final course score with mean of 0 and standard deviation of 1 (referred to as final course z-score in the rest of the paper), $\epsilon_i$ represents the "noise term" that accounts for all other effects not explained by the interaction states on the modules. We assume that $\epsilon_i$ are identical and independently normally distributed with mean 0.

For each of the three levels, the reference state is set to be the first state with $s = 1$. According to the three levels, we study the effects with number of states $S$ to be 3, 6 and, 9 respectively. Specifically, the reference state is listed as follows for each level.

I. Final Course z-score~ 3 states. Reference State: P
II. Final Course z-score~ 6 states. Reference State: BSP
III. Final Course z-score~ 9 states. Reference State: BSPN

In each level, the reference state is selected as the interaction state that is most likely associated with the highest level of content knowledge from an instructor's point of view. The intercept reflects the predicted final course z-score if all modules are in the reference state.

For comparison, we also create a baseline linear regression model between the number of modules a student failed and aborted and their final course score:

$$y_i = \alpha_0 + \alpha_F x_{F,i} + \alpha_A x_{A,i} + \epsilon_i \quad (2)$$

where $y_i$ is the standardized final course z-score for student $i$ and $x_{F,i}$ and $x_{A,i}$ are the number of modules the student failed or aborted, respectively. The parameter $\alpha_0$ stands for the

$$y_i = \beta_0 + \sum_{s=2}^{S} \beta_{1,s}\delta_{i,1,s} + \sum_{s=2}^{S} \beta_{2,s}\delta_{i,2,s} + \cdots + \sum_{s=2}^{S} \beta_{10,s}\delta_{i,10,s} + \epsilon_i \quad (1)$$

In the model above, $\delta_{i,m,s}$ are dummy variables with $\delta_{i,m,s} = 1$ if the $i$th student has interaction state $s$ for module $m$, and $\delta_{i,m,s} = 0$ otherwise, for $i = 1,2,\ldots,n, m = 1,2,\ldots,10$ and $s = 1,2,\ldots,S$. Here $n = 223$ is the number of students who completed all 10 modules, and $S$ is the maximum number of interaction states defined in each level of linear model. The variables $\delta_{i,m,s}$ combine information contained in the module number, such as different content and difficulty of each module, with students' interaction states. Consequently, the model parameter $\beta_0$ represents the expected final course score for students in a "reference state" for every module, while $\beta_{m,s}$ measures the difference in the final score by being in state $s$ in module $m$ compared to the reference state.

expected score of students who passed all modules and $\alpha_F$ (and $\alpha_A$, respectively) represents the amount of points decreased in the course final z-score for failing (aborting, respectively) one more module.

### 2. Addressing Collinearity within regression variables using LASSO

**Collinearity and Regularized Regression:** To construct the linear model (1), we are estimating $p = 10S - 9$ unknown coefficients, i.e., 21, 51, and 81 respectively for level I, II, and III. The fact that $p$ is nonnegligible to the number of students $n = 207$ can induce significant issue in the regression. In particular, it is likely that the space constructed by the predictors is (nearly) singular, which means some of the covariates are (nearly) linear combinations of others. This issue is known as collinearity and it can cause a highly inaccurate and unstable, if not non-existent, model estimation since the ordinary least square solution of 1 relies on the assumption that the covariate space is nonsingular. In presence of collinearity, the estimated relationship can be spurious and redundant, as the effect of one covariate can be replaced by the combination of others.

In remedy of the collinearity, we employ LASSO (Least Absolute Shrinkage and Selection Operator) estimation [29, 42], assuming only a small proportion of the states significantly influence the final course score. The LASSO regression regularizes the estimation by imposing a penalty of model size to the square sum of errors, defined as in the following equation:

$$(\hat{\beta}_0, \hat{\boldsymbol{\beta}}) = \operatorname{argmin}_{(\beta_0, \boldsymbol{\beta})} \sum_{i=1}^{n} (y_i - \beta_0 - \boldsymbol{\delta}_i^\top \boldsymbol{\beta})^2 + \lambda \|\boldsymbol{\beta}\|_1, \qquad (3)$$

where the vector $\boldsymbol{\delta}_i = (\delta_{i,m,s}, 1 \leq m \leq 10, 1 \leq s \leq S)^\top$ contains the binary state dummy variable for each state in all ten modules, and $\boldsymbol{\beta}$ contains the corresponding coefficients. The tuning parameter $\lambda$ controls the strength of penalty in the model, and hence the sparsity of the estimation. We select $\lambda$ by a 10-fold cross validation with the minimum mean squared error. LASSO estimation assumes that a small subset of $\boldsymbol{\beta}$ is nonzero and is well-known for its model selection consistency under certain conditions (c.f., [44]). In other words, the estimator (3) is able to select the correct subset of features relevant to the overall course score with high probability. That is, with a large sample size, model (3) selects the relevant models and states and excludes the irrelevant with probability near one. We use for feature selection and then regress the final course z-score against the selected modules and states. Assume only a subset of interaction states in all the modules are relevant to students' final performance in the course, denoted as $S_0 = \{1 \leq j \leq p : \beta_j \neq 0\}$. Let $\hat{S}_0 = \{1 \leq j \leq p : \hat{\beta}_j \neq 0\}$ be the index set of significant features selected in equation (3) and $\boldsymbol{\delta}_{\hat{S}_0}$ be the design matrix for the corresponding modules and states. We estimate the corresponding coefficients $\boldsymbol{\beta}^\star$ from the following regression:

$$\boldsymbol{y} = \boldsymbol{\delta}_{\hat{S}_0}^\top \boldsymbol{\beta}^\star + \boldsymbol{\epsilon}, \qquad (4)$$

where $\boldsymbol{y}$ and $\boldsymbol{\epsilon}$ are the vector form of the final course z-score and noise respectively. Note in equation (4), the irrelevant states are not included in the predictors.

### D. Visualizing students' transition between interaction states in an OLM sequence

To visualize how students transition between interaction states from one module to the next, we plot data from the 10 modules on a sequence of nine parallel coordinate diagrams, as shown in Figure 3, Figure 4, and Figure 5. The two vertical axes on each of the nine diagrams represent the interaction states on two adjacent modules. Each student is represented as a line starting from one interaction state on the left axis and ending on another interaction state on the right axis. One or more overlapping lines form a path indicating a transition between two interaction states on two adjacent modules, where a horizontal path means that one or more student remained in the same state on the two modules. The student population is divided equally into top 1/3, middle 1/3 and bottom 1/3 cohorts according to their final course score, with each cohort plotted on its own sequence of parallel coordinate diagrams. The most populated major paths that add up to half of the population within each cohort are highlighted by a yellow line, with the line widths proportional to the size of the major path. The current visualization scheme has two differences from the version in the earlier study [31]: 1. The ordering of states is now based on the results of the linear model. States that are more frequently correlated with lower course grades are placed lower on the graph, with reference state being placed at the top of the graph. 2. Adjacent paths are no longer clustered into a single path, as it cannot be argued that adjacent states are more similar to each other than distant states.

In addition, variables in the linear model selected by the LASSO estimation algorithm are highlighted by three types of labels on the axis: hollow triangles represent selected variables with $\beta^{\star}_{\hat{S}_0}$ not significantly different from zero, solid squares represent variables with $\beta^{\star}_{\hat{S}_0}$ significantly different from 0 at $\alpha < 0.05$ level, and solid spheres represent with $\beta^{\star}_{\hat{S}_0}$ significantly different from 0 at $\alpha < 0.01$ level. Selected variables with $\beta^{\star}_{\hat{S}_0} > 0$ are represented by Dark Cyan (#1A9F76) labels, and those with $\beta^{\star}_{\hat{S}_0} < 0$ are represented by Pollo Blue (#8DA0CB) labels. Each label is repeated three times on the three graphs for the three cohorts.

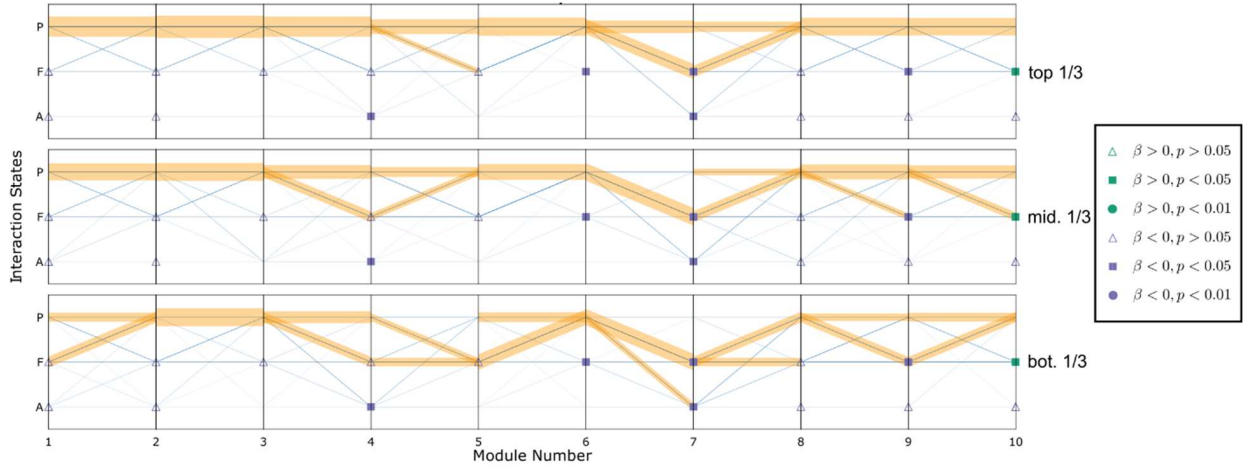

*Figure 3: Parallel coordinate graphs using Level I (three) states.*

## IV. Results

### A. Baseline Model

The intercept and coefficients of the baseline regression model (Adjusted $R^2 = 0.18$, $F = 16.21$, $p < 0.01$) are listed in Table 3. As expected, the average final score for students who passed all modules differs significantly from the average of all students, and the number of both failed and aborted modules are negatively correlated with final course score, with correlation coefficients significantly different from zero at $\alpha < 0.01$ level.

*Table 3: Estimated coefficients and p-values for the baseline linear regression model (2).*

| State | Coeff.($\alpha$) | p |
|---|---|---|
| Intercept | 0.74 | 0.00** |
| F | -0.15 | 0.00** |
| A | -0.32 | 0.00** |

### B. Level I: Three Interaction States

For level I (three states on each module), 17 out of 21 variables are selected by the LASSO algorithm, resulting in a linear model of Adjusted $R^2 = 0.20$, $F = 4.07$, $df = 189$, $p < 0.01$. The coefficients of the model are shown in Table 4. Most of the variables are negatively correlated with the final score, which is expected since the P state is selected as the reference state for each module. In addition to the intercept, six variables have coefficients that are significantly different from zero, in which five of those are on modules 6-10. This is likely because the difficulty of the modules increases towards the end of the sequence. Surprisingly, the F state on module 10 is positively correlated with the final score, indicating that students with high final scores are more likely to fail on this module.

*Table 4: Estimated coefficients and p-values of regression (4) for the three-state model (Level I).*

| Module | State | Coeff.($\beta^{\star}$) | p |
|---|---|---|---|
| NA | Intercept | 0.80 | 0.00** |
| m1 | A | -0.14 | 0.61 |
| m1 | F | -0.24 | 0.08 |
| m2 | A | -0.31 | 0.49 |
| m2 | F | -0.13 | 0.36 |
| m3 | F | -0.06 | 0.72 |
| m4 | A | -0.50 | 0.03* |
| m4 | F | -0.17 | 0.18 |
| m5 | F | -0.21 | 0.09 |
| m6 | F | -0.40 | 0.04* |
| m7 | A | -0.48 | 0.01* |
| m7 | F | -0.32 | 0.03* |
| m8 | A | -0.36 | 0.29 |
| m8 | F | -0.17 | 0.19 |
| m9 | A | -0.52 | 0.24 |
| m9 | F | -0.30 | 0.02* |
| m10 | A | -1.02 | 0.23 |
| m10 | F | 0.25 | 0.04* |

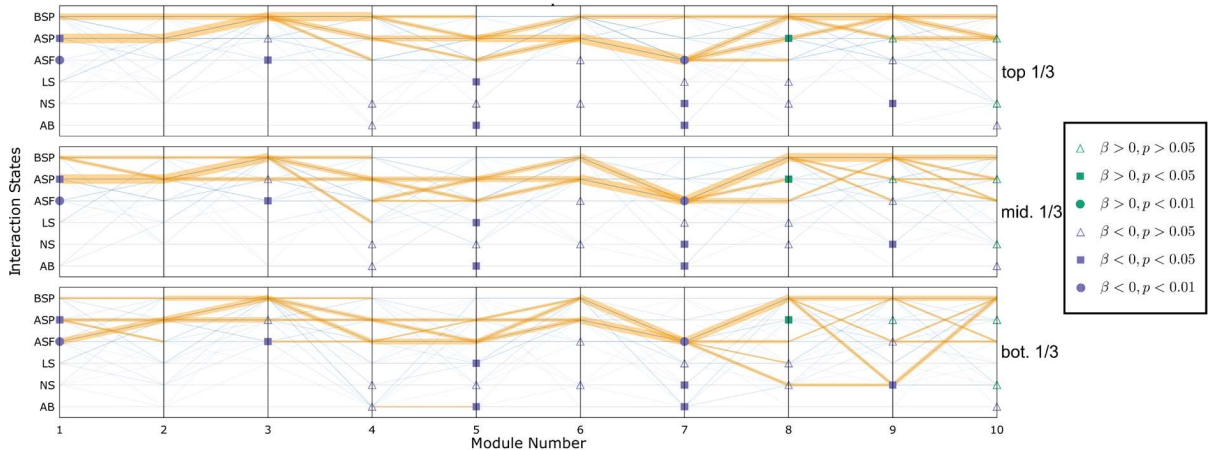

*Figure 4: Parallel coordinate graphs using Level II (six) states.*

On the parallel coordinate graph (Figure 3), the three states are ordered as P, F, A, since on all modules (except on m10) the coefficients for both F and A states are negative, with the A states being more negative. Four of the six significant variables correspond to the start or end point of a major path. Of which, m7-A is on the end of a significant path in the bottom cohort only; m7-F is at the junction of two major paths on all three cohorts; m9-F is on 2 major paths in the bottom cohort and one major path in the middle cohort; m10-F is on the end of a major path in the middle cohort only.

### C. Level II: Six Interaction States

For level II , 24 out of 51 variables are selected by the LASSO algorithm, resulting in a linear model with adjusted $R^2 = 0.33$, $F = 5.268$, $df = 182$, $p < 0.01$, the coefficients of which are shown in Table 5. In addition to the intercept, 10 variables have coefficients that are significantly different from zero at the $\alpha = 0.05$ level, one of which, m1-ASF, is significant at the $\alpha = 0.01$ level. Most of the variables are negatively correlated with the final score, except for m8-ASP, m9-ASP, m10-ASP, and m10-NS.

The ordering of states on the corresponding parallel coordinate graph (Figure 4) reflects the fact that LS, NS and AB states on multiple modules are significantly negatively correlated with final course score. Of the 10 significant variables, 3 of which: m5-LS, m7-NS, m7-AB are not located on any major path in any of the cohorts, and one variable, m5-AB, is located on a small major path in the bottom third cohort. It is likely that those variables reflect the behavior of a small fraction of students with exceptionally low final course score.

*Table 5: Estimated coefficients and p-values of regression (4) for the six-state model (Level II).*

| Module | State | Coeff.($\beta^\star$) | p |
|---|---|---|---|
| NA | Intercept | 0.75 | 0.00** |
| m1 | ASP | -0.26 | 0.04* |
| m1 | ASF | -0.51 | 0.00** |
| m3 | ASP | -0.24 | 0.06 |
| m3 | ASF | -0.40 | 0.02* |
| m4 | NS | -0.55 | 0.10 |
| m4 | AB | -0.36 | 0.16 |
| m5 | LS | -0.45 | 0.04* |
| m5 | NS | -0.21 | 0.43 |
| m5 | AB | -1.18 | 0.02* |
| m6 | ASF | -0.19 | 0.41 |
| m6 | NS | -0.51 | 0.11 |
| m7 | ASF | -0.36 | 0.01* |
| m7 | LS | -0.51 | 0.06 |
| m7 | NS | -0.79 | 0.03* |
| m7 | AB | -0.57 | 0.02* |
| m8 | ASP | 0.31 | 0.03* |
| m8 | LS | -0.22 | 0.33 |
| m8 | NS | -0.13 | 0.54 |
| m9 | ASP | 0.11 | 0.42 |
| m9 | ASF | -0.14 | 0.35 |
| m9 | NS | -0.46 | 0.02* |
| m10 | ASP | 0.25 | 0.06 |
| m10 | NS | 0.29 | 0.11 |
| m10 | AB | -1.25 | 0.11 |

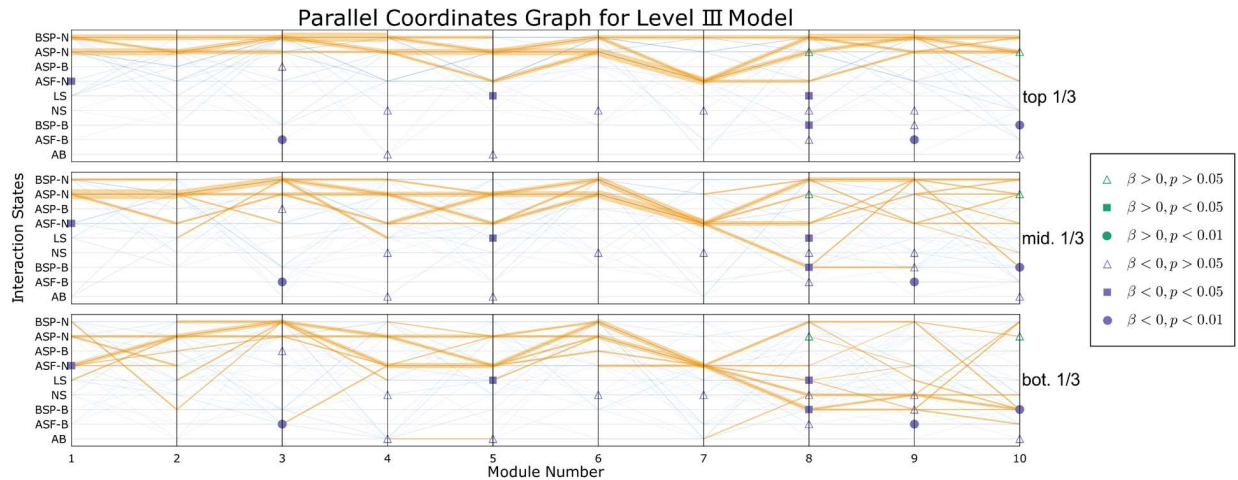

Figure 5: Parallel coordinate graphs using Level III (nine) states.

Of the remaining 6 significant variables that are also located on at least one major path, there are several noteworthy observations:

1. Among passing states, ASP and BSP (reference state) are similar in their correlation with final course score, except on m1 and m8. On m1, ASP is significantly negatively correlated with final course score, but is also on a major path in all three cohorts. A possible explanation is that the fraction of students with the highest final scores can pass this module, which introduces the definition of kinetic energy, prior to studying the content. Surprisingly, m8-ASP is positively correlated with final course grade compared to m8-BSP, and is on a major path in both the top and middle cohort, but not in the bottom cohort. This implies that students with high final scores are more likely to pass the module after studying the IC rather than passing on their initial attempt. Both m9-ASP and m10-ASP are also positively correlated with final score, with m10-ASP being marginally significant ($p = 0.06$).

2. For failing states (ASF, LS, and NS), m7-ASF still sits on multiple major paths on all three cohorts, suggesting that only a few top students in the class can pass m7, and the IC of m7 not helping the majority of students. On the other hand, m9-NS is connects three major paths in the bottom cohort, while almost no student occupied that state.

### D. Level III: Nine Interaction States

For level III, 20 out of 90 variables are selected by the LASSO algorithm (Table 6), producing a minimum linear model with adjusted $R^2 = 0.39, F = 7.69, df = 186, p < 0.01$. 8 of the 20 variables have correlation coefficients that are significantly different from 0 at the $\alpha = 0.05$ level, two of which are significant at the $\alpha = 0.01$ level. Two variables have positive correlation coefficients, but neither are significant.

For the parallel coordinates graph, we noticed that m3-ASF-B, m9-ASF-B and m10-BSP-B are the only three variables that are significantly correlated with lower final score at the $\alpha = 0.01$ level, and both ASF-B and BSP-B states also have significant negative correlations on several other modules. In comparison, BSP-N (reference state) is positively correlated with final score (significant positive intercept), while ASFN on most modules are indistinguishable from BSPN since it is not selected by LASSO on any module except m1. To visually represent this large difference between ASF-B/BSP-B and ASF-N/BSP-N, we placed BSP-B and ASF-B at the bottom of the graph just above AB, where ASP-B is placed next to ASP-N since our algorithm did not detect any difference between the two states. The other states are ordered similar to level II.

When compared to the level II graph, major paths and LASSO selected variables for m1-m6 are quite similar, indicating that on those modules, most BSP, ASP and ASF events in level II belong to BSP-N, ASP-N and ASF-N in level III. The two noteworthy features are: 1. while m1-ASP was significantly correlated with final score in level II, m1-ASP-N and m1-ASP-B are not selected by LASSO as necessary variables in level III; 2. m3-ASF-B is a significant negatively correlated variable, similar to m3-ASF in level II. On the other hand, the level III model tells a very different story on m7-m10:

1. Most interaction states on m7 are no longer selected by LASSO for explaining the variance in the final course grade. Compared to level II, in which 4 states are selected with 3 being significant, only m7-NS is selected in level III, and the correlation is not significant. Meanwhile, m7-ASF-N still serves as a "hub" connecting multiple major paths in all three cohorts.
2. BSP-B and BSP-N sates on m8-m9 have different compositions between the three cohorts. On m8-m9, most BSP events in the top cohort belongs to BSP-N, while for the bottom cohort a significant fraction of BSP events belong to BSP-B. This seems to be a likely reason why m8-ASP-N in level III has a much weaker positive correlation compared to what was observed for m8-ASP in level II, since the current reference state, BSP-N, is occupied by more students with higher course score.
3. Interaction states on m8-m10 differ significantly between top and bottom cohorts. With the current arrangement of states, the bottom third cohort aggregated onto a "corridor" consisting of major paths between LS, NS, BSP-B, and ASF-B states extending from m8 to m10, "anchored" by several significant LASSO selected variables. In contrast, this corridor is almost empty for the top cohort, and less populated for the middle third cohort. The top cohort is mostly concentrated on BSP-N and ASP-N states between m8-m10, which are only sparsely occupied by the bottom cohort.

## V. Discussion

### A. Including more contextual information led to better descriptions of student behavior

Our analysis demonstrates that by increasing the amount of contextual information associated with each pass-fail event, we can obtain more informative and accurate descriptions of students' online learning behavior.

The baseline regression model (equation (2)) reveals little more than the fact that high performing students pass more modules. By including the module number information, the level I model shows that passing modules m6-m9 are better indicators of higher final course score. However, it is difficult to understand why failing on m10 is positively correlated with final course score. Note that the LASSO algorithm selected 17 out of 20 variables in this model, indicating that it has only limited ability to identify characteristic behavioral differences between students with high and low total course score. The $R^2$ value for both models (0.18 and 0.22 respectively) are below the recommended criteria of 0.25 for moderate effect in social science data [45], whereas the level II (0.33) and level III (0.39) models are within the range of moderate effects ($0.25 < R^2 < 0.64$).

Table 6: Estimated coefficients and p-values of regression (4) for the nine-state model (Level III).

| Module | State | Coeff.($\beta^\star$) | p |
|---|---|---|---|
| NA | Intercept | 0.52 | 0.00** |
| m1 | ASF-N | -0.37 | 0.01* |
| m3 | ASF-B | -0.63 | 0.01** |
| m3 | ASP-B | -0.70 | 0.12 |
| m4 | NS | -0.53 | 0.08 |
| m4 | AB | -0.17 | 0.48 |
| m5 | LS | -0.45 | 0.02* |
| m5 | AB | -0.98 | 0.05 |
| m6 | NS | -0.33 | 0.28 |
| m7 | NS | -0.45 | 0.16 |
| m8 | ASP-N | 0.09 | 0.55 |
| m8 | ASF-B | -0.55 | 0.11 |
| m8 | LS | -0.50 | 0.02* |
| m8 | BSP-B | -0.33 | 0.05* |
| m8 | NS | -0.24 | 0.26 |
| m9 | ASF-B | -1.13 | 0.00** |
| m9 | BSP-B | -0.03 | 0.85 |
| m9 | NS | -0.21 | 0.26 |
| m10 | ASP-N | 0.19 | 0.16 |
| m10 | BSP-B | -0.65 | 0.00** |
| m10 | AB | -1.35 | 0.07 |

The level II states added the contextual information on whether each pass-fail event happened before or after accessing the instructional materials. The level II model reveals that on modules m5, m7, m8 and m9, students with lower final score not only have lower passing rates, but are also more reluctant to access the instructional materials after repeated failure (LS and NS states). This could imply that those students either have less motivation to study or have otherwise lost confidence in their ability to learn from the IC. On the other hand, two observations are difficult to make sense of. First, the ASP states on m8, m9 and m10 are positively associated with final score, which implies that students with higher scores are more likely to fail their initial attempts and needed to study the IC. Second, Figure 4 shows that many students in the bottom third cohort transitioned from NS and ASF states on m9 to BSP state on m10, which contains a harder problem than m9 in the AC.

The level III model included information on whether the pass-fail event was completed over a brief interval (less than 35 seconds). The addition of this information seems to be important for identifying characteristic behavioral differences between students with high and low final course scores, as it allows the LASSO algorithm to select only 20 out of 90 variables. The resulting model accounted for more variance in the final course score using 4 fewer variables than the level II model.

The level III parallel coordinate graph (Figure 5) shows a clear "corridor" from m8 to m10 for the bottom third cohort, consisting of major paths connecting either brief passing attempts (BSP-B) or consecutive failed attempts without study (LS or NS). In contrast, the top third cohort mainly concentrated on normal passing attempts either before or after studying the IC (BSP-N and ASP-N) on the same modules, whereas the middle third cohort has more failed normal attempts (ASF-N). Remarkably, for all three cohorts, the major paths between m8-m10 all originated from the same ASF-N state on m7. This observation suggests that failing on m7 is not a characteristic difference between high and low performing students, but their different choices after experiencing the setback on m7 is: while the top and most of the middle cohort continued with learning (with the middle cohort being less successful), most of the bottom cohort gave up and resorted to guessing on the following modules.

The level III model also provides an explanation for the anomalous observations on level I and II models: many P and BSP events from the bottom 1/3 cohort on m9 and m10 are BSP-B events (attempts shorter than 35 seconds), while only a few students in this cohort studied the IC of the module. Based on previous research [18, 23], one possible interpretation is that students in the bottom 1/3 cohort are more likely to have copied the answers to the problems from another source.

### B. Implications for Instructors

One of the important goals of learning analytics is to provide instructors with actionable recommendations to improve student learning. In that regard, the level III model is far superior to the other models.

The simple baseline model and level I model both rely on pass-fail events alone, which is similar to what is provided by many commercial online homework platforms. According to these two models, the average instructor can do little more than ask students to "work harder and pass more modules, especially on m6-m9". The Level II model suggests that some students might have lost confidence toward the end, but the patterns are inconsistent. In addition, Level I and II models could mislead the instructor into believing that the bottom third cohort eventually mastered the content or even outperformed the top and middle cohorts on m9 and m10.

On the other hand, the level III model tells a more complete and accurate story with three main takeaways:

1. On modules m1-m6, there are no qualitative differences in learning strategy for students with varying levels of ability to succeed in the course. In other words, almost everyone is trying to learn in the beginning.
2. Module 7 is challenging for most students as the instructional materials are insufficient for helping them learning how to solve the problems in the AC.
3. After experiencing a setback on m7, students with low course final scores are much more likely to employ a guessing (or copying) strategy on the rest of the modules.

Given those takeaways, rather than telling students to "study harder" or "do better," a more helpful message could be "Everybody experiences setbacks – it is alright to fail! The key to success is to not give up." In addition, two interventions could potentially be beneficial for boosting students' confidence:
1. Improve the quality of instruction on m7 to increase the chance of success especially for low performing students.
2. Conduct activities that develop a growth mindset, which has been shown to be beneficial for student success [46–48].

Looking at the content of each module (which can be accessed via [37]), m1-m6 mostly focused on introducing the basic concepts of work and mechanical energy, whereas m7-m10 were designed to develop students' ability to solve numerical problems. The transition from conceptual understanding to mathematical modeling took place between m6, which contains two conceptual problems on the conservation of mechanical energy, and m7, which contains both a conceptual problem and a numerical calculation problem on the same topic. Our resultss suggest that this transition is very challenging for most students, and could have an impact on the confidence of some students. Therefore, instructors need to provide more scaffolding to facilitate students in this transition. A valuable future direction will be to investigate if the difficulty in the coneptual-mathematical transition can be observed for other topics in introductory physics and in other learning environments.

### C. Implications for researchers conducting data-driven online learning research

First of all, we demonstrated that instead of employing more sophisticated algorithms, fine tuning different parameters, or using larger data sets, including detailed contextual information for each event analyzed can in some cases also be an effective approach for improving not only the accuracy of data analysis models, but more importantly in improving the ability to provide actionable and targeted instructional suggestions for instructors.

Second, this study highlights the importance of the instructional design and platform capability in learning analytics. The contextual data that are crucial for the construction of the level II and III models are grounded in the unique OLM design blending assessment with instructional resources, which is made possible by the flexibility of the Obojobo platform. It is often the case that platform capability and instructional design can determine both the variety and accuracy of information that can be extracted from student log data [22, 48], and in turn limits the depth of learning analytics. For example, the RISE project [50] is limited to simple analysis and visualization with limited contextual information, using data from generic online learning platforms. Therefore, it can be beneficial for all parties involved if data scientists and online learning researchers play a more actively role in the design, development or adoption of online learning platforms and online courses, rather than passively stay on the receiving end of educational data.

### D. Caveats

One limitation of the current analysis is the use of a universal 35 seconds cutoff between Brief and Normal attempts. While this stringent criterion is favorable for avoiding false positives, it may not capture a significant number students who are not trying very hard on complex calculation problems that cannot be correctly solved within several minutes even by experts. This might explain why we still observe some students in the bottom cohort shift from Late Study and Abort states on m9 to the BSP-N state on m10. In fact, for m9 and m10, exploratory data analysis [31] identified a separate distribution spending longer than average time solving the problem, while achieving a better correct response rate. Spending more than average time on those problems could be a characteristic behavior of the top 1/3 cohort just as "Brief" problem solving is characteristic for the bottom 1/3 cohort.

Another imperfection of the current analysis is that the scores on the OLM sequence are included in the total final course grade, which violates the conditions for linear regression. However, we think that this is a negligible small effect because: 1. the OLM sequence only accounts for 9% of the total grade and 2. all students received at least 90% of the score if they passed the module in 5 attempts. As a result, the failed states used in the linear model does not directly correlate to the module scores.

## E. General Discussion and Future Directions

It is important to clarify that the purpose of the current work is not to create a predictive model for the course final score. Instead, our focus is on identifying and making sense of different behavior patterns among students with different levels of course performance, as well as demonstrating the value of integrating contextual information with events to obtain a more accurate and interpretable description of student learning. This choice of focus provides justification for a number of decisions made in the current study.

First, we did not use one part of our data to generate the regression model and reserve other parts for verification, as would be the standard process for creating a predictive model. Such an operation is not essential for identifying and understanding students' online learning behavior. Another reason is that not enough data was collected at the time the analysis was conducted.

Second, we chose the total final course score as the dependent variable because it is the most straightforward and generic way to classify high, middle, and low performing students in a class, and is most suitable for answering our research question. Student scores on a single assessment, or on part of an assessment related to the topic of the module would be more suitable for a predictive model.

Third, we chose to not include several types of available data such as the time of each submission relative to the due date, the number of practice problems solved during learning, or the demographics of the student population. All of which could have improved the predictive power of the model, but would not answer our research question.

Although not a predictive model itself, the current study is an essential first step towards creating better future predictive models. Existing predictive models are successful at identifying at-risk students with high accuracy, but often have only limited ability to provide specific and useful recommendation for both students and instructors. For example, students identified to be at-risk by the Course Signal program receive little more than e-mail and text messages alerting them of their status [9]. The current study demonstrated the possibility of overcoming such shortage by collecting and integrating contextual information with individual learning events.

Another important question that the current study lays the groundwork for answering is how the design of online learning resources may shape students' learning behavior and learning outcomes. An actionable next step along this direction is to examine whether improvements recommended by the level III model could lead to detectable changes in students' enagagement pattern.

Futhremore, the OLMs' unique design allows for de-facto pre and post tests to be conducted on each module [41], providing researchers with a new tool to measure students' learning gains at much higher frequency than existing methods. This could lead to new insight into the relation between students' learning behavior and learning outcomes in an online environment. Much future work is needed to either develop new analysis tools, or adopt similar existing methods [51] to properly measure learning gain from OLM data.

Finally, a more general question is whether including contextual information could benfet the analysis of other types of data commonly studied in the field of PER. For example, we may be able to gain new insight into students' response data from standard assessment instruments, such as the Force Concept Inventory, by studying students' response time on each question, or considering the level to which classroom instruction is aligned with the test questions, using analysis methods similar to those developed in the current study.

# VI. Acknowledgement

The authors would like to thank the Learning Systems and Technology team at UCF for developing the Obojobo platform. This research is partly supported by NSF Award No. DUE-1845436 and the Advancement of Early Career Researchers (AECR) Program at the University of Central Florida.

# VII. References

[1] G. Kortemeyer, Work Habits of Students in Traditional and Online Sections of an Introductory Physics Course: A Case Study, J. Sci. Educ. Technol. **25**, 697 (2016).

[2] G. Kortemeyer, Correlations between Student Discussion Behavior, Attitudes, and Learning, Phys. Rev. Spec. Top. - Phys. Educ. Res. **3**, 1 (2007).

[3] M. Formanek, S. Buxner, C. Impey, and M. Wenger, Relationship between Learners ' Motivation and Course Engagement in an Astronomy Massive Open Online Course, Phys. Rev. Phys. Educ. Res. **15**, 20140 (2019).

[4] S. Y. Lin, J. M. Aiken, D. T. Seaton, S. S. Douglas, E. F. Greco, B. D. Thoms, and M. F. Schatz, Exploring Physics Students' Engagement with Online Instructional Videos in an Introductory Mechanics Course, Phys. Rev. Phys. Educ. Res. **13**, 1 (2017).

[5] J. Qiu, J. Tang, T. X. Liu, J. Gong, C. Zhang, Q. Zhang, and Y. Xue, Modeling and Predicting Learning Behavior in MOOCs, in *Proc. Ninth ACM Int. Conf. Web Search Data Min. - WSDM '16* (ACM Press, New York, New York, USA, 2016), pp. 93–102.

[6] R. F. Kizilcec, C. Piech, and E. Schneider, Deconstructing Disengagement : Analyzing Learner Subpopulations in Massive Open Online Courses, Lak '13 10 (2013).

[7] I. Borrella, S. Caballero-Caballero, and E. Ponce-Cueto, Predict and Intervene: Addressing the Dropout Problem in a MOOC-Based Program, in *Proc. Sixth ACM Conf. Learn. @ Scale - L@S '19* (ACM Press, New York, New York, USA, 2019), pp. 1–9.

[8] R. Martinez-Maldonado, K. Yacef, J. Kay, A. Kharrufa, and A. Al-Qaraghuli, Analysing Frequent Sequential Patterns of Collaborative Learning Activity Around an Interactive Tabletop., in *EDM 2011 - Proc. 4th Int. Conf. Educ. Data Min.* (2011), pp. 111–120.

[9] K. E. Arnold, M. D. Pistilli, and K. E. Arnold, Course Signals at Purdue: Using Learning Analytics to Increase Student Success, 2nd Int. Conf. Learn. Anal. Knowl. 2 (2012).

[10] J. W. You, Identifying Significant Indicators Using LMS Data to Predict Course Achievement in Online Learning, Internet High. Educ. **29**, 23 (2016).

[11] Y. Li, C. Fu, and Y. Zhang, When and Who at Risk ? Call Back at These Critical Points, in *Proc. 10th Int. Conf. Educ. Data Min.* (2017), pp. 168–173.

[12] S. M. Jayaprakash, E. W. Moody, E. J. M. Lauria, J. R. Regan, and J. D. Baron, Early Alert of Academically At-Risk Students: An Open Source Analytics Initiative, J. Learn. Anal. **1**, 6 (2014).

[13] R. S. Baker, D. Lindrum, M. J. Lindrum, and D. Perkowski, Analyzing Early At-Risk Factors in Higher Education e- Learning Courses, Proc. 8th Int. Conf. Educ. Data Min. 150 (2015).

[14] D. Gašević, S. Dawson, and G. Siemens, Let ' s Not Forget : Learning Analytics Are about Learning, TechTrends64, **59**  (2015).

[15] P. H. Winne and A. F. Hadwin, Studying as Self-Regulated Learning, (2017).

[16] B. R. Wilcox and S. J. Pollock, Investigating Students ' Behavior and Performance in Online Conceptual Assessment, Phys. Rev. Phys. Educ. Res. **15**, 20145 (2019).

[17] D. T. Seaton, G. Kortemeyer, Y. Bergner, S. Rayyan, and D. E. Pritchard, Analyzing the Impact of

Course Structure on Electronic Textbook Use in Blended Introductory Physics Courses, Am. J. Phys. **82**, 1186 (2014).

[18] G. Alexandron, J. A. Ruiperez-Valiente, Z. Chen, Pedro J. Muñoz-Merino, and D. E. Pritchard, Copying @ Scale: Using Harvesting Accounts for Collecting Correct Answers in a MOOC, Comput. Educ. **108**, (2017).

[19] D. J. Palazzo, Y. J. Lee, R. Warnakulasooriya, and D. E. Pritchard, Patterns, Correlates, and Reduction of Homework Copying, Phys. Rev. Spec. Top. - Phys. Educ. Res. **6**, (2010).

[20] D. T. Seaton, Y. Bergner, I. Chuang, P. Mitros, and D. E. Pritchard, Who Does What in a Massive Open Online Course?, Commun. ACM **57**, 58 (2014).

[21] B. Gutmann, G. E. Gladding, M. Lundsgaard, and T. Stelzer, Mastery-Style Homework Exercises in Introductory Physics Courses: Implementation Matters, Phys. Rev. Phys. Educ. Res. (n.d.).

[22] M. W. Guthrie and Z. Chen, Comparing Student Behavior in Mastery and Conventional Style Online Physics Homework, in *Phys. Educ. Res. Conf. Proc. 2019*, edited by Y. Cao, S. Wolf, and M. B. Bennett (Provo, UT, 2019).

[23] Z. Chen, S. Lee, and G. Garrido, Re-Designing the Structure of Online Courses to Empower Educational Data Mining, in *Proc. 11th Int. Educ. Data Min. Conf.*, edited by K. Elizabeth Boyer and M. Yudelson (Buffalo, NY, 2018), pp. 390–396.

[24] G. Alexandron, J. A. Ruipérez-Valiente, Z. Chen, P. J. Muñoz-Merino, and D. E. Pritchard, Copying@Scale: Using Harvesting Accounts for Collecting Correct Answers in a MOOC, Comput. Educ. **108**, (2017).

[25] R. Warnakulasooriya, D. J. Palazzo, and D. E. Pritchard, Time to Completion of Web-Based Physics Problems with Tutoring., J. Exp. Anal. Behav. **88**, 103 (2007).

[26] C. L. Barry, S. J. Horst, S. J. Finney, A. R. Brown, and J. P. Kopp, Do Examinees Have Similar Test-Taking Effort? A High-Stakes Question for Low-Stakes Testing, Int. J. Test. **10**, 342 (2010).

[27] S. L. Wise and X. Kong, Response Time Effort: A New Measure of Examinee Motivation in Computer-Based Tests, Appl. Meas. Educ. **18**, 163 (2005).

[28] S. L. Wise, D. A. Pastor, and X. J. Kong, Correlates of Rapid-Guessing Behavior in Low-Stakes Testing: Implications for Test Development and Measurement Practice, Appl. Meas. Educ. **22**, 185 (2009).

[29] E. J. Theobald, M. Aikens, S. Eddy, and H. Jordt, Beyond Linear Regression: A Reference for Analyzing Common Data Types in Discipline Based Education Research, Phys. Rev. Phys. Educ. Res. **15**, 20110 (2019).

[30] R. Tibshirani, Regression Shrinkage and Selection via the Lasso, J. R. Stat. Soc. Ser. B **58**, 567 (1996).

[31] G. Garrido, M. W. Guthrie, and Z. Chen, How Are Students ' Online Learning Behavior Related to Their Course Outcomes in an Introductory Physics Course ?, in *Phys. Educ. Res. Conf. 2019*, edited by Y. Cao, S. Wolf, and M. B. Bennett (Provo, UT, 2019).

[32] Center for Distributed Learning, Obojobo, https://next.obojobo.ucf.edu/.

[33] J. W. Morphew, G. E. Gladding, and J. P. Mestre, Effect of Presentation Style and Problem-Solving Attempts on Metacognition and Learning from Solution Videos, Phys. Rev. Phys. Educ. Res. **16**,

10104 (2020).

[34] D. Schwartz, J. Bransford, and D. Sears, Efficiency and Innovation in Transfer, Transf. Learn. from a … 1 (2005).

[35] W. Fakcharoenphol, E. Potter, and T. Stelzer, What Students Learn When Studying Physics Practice Exam Problems, Phys. Rev. Spec. Top. - Phys. Educ. Res. **7**, 1 (2011).

[36] C. Singh and D. Rosengrant, Multiple-Choice Test of Energy and Momentum Concepts, Am. J. Phys. **71**, 607 (2003).

[37] Z. Chen, Obojobo Sample Canvas Course, https://canvas.instructure.com/courses/1726856.

[38] H. Wickham, Tidyverse: Easily Install and Load the "Tidyverse," (2017).

[39] R Core Team, R: A Language and Environment for Statistical Computing, Doc. Free. Available Internet Http//Www. r-Project. Org (2019).

[40] D. L. Schnipke and D. J. Scrams, Modeling Item Response Times with a Two-State Mixture Model- a New Approach to Measuring Speededness, J. Educ. Meas. **34**, 213 (1999).

[41] Z. Chen, G. Garrido, Z. Berry, I. Turgeon, and F. Yonekura, Designing Online Learning Modules to Conduct Pre- and Post-Testing at High Frequency, in *2017 Phys. Educ. Res. Conf. Proc.* (American Association of Physics Teachers, Cincinnati, OH, 2018), pp. 84–87.

[42] M. O. Prates and C. R. B. Cabral, Mixsmsn: Fitting Finite Mixture of Scale Mixture of Skew-Normal Distributions Marcos, J. Stat. Softw. **30**, 1 (2009).

[43] J. Friedman, T. Hastie, and R. Tibshirani, Regularization Paths for Generalized Linear Models via Coordinate Descent, J. Stat. Softw. **33**, 1 (2010).

[44] P. Zhao and Y. Bin, On Model Selection Consistency of Lasso, J. Mach. Learn. Res. **7**, 2541 (2006).

[45] C. J. Ferguson, An Effect Size Primer: A Guide for Clinicians and Researchers, Prof. Psychol. Res. Pract. **40**, 532 (2009).

[46] D. S. Yeager, D. Paunesku, G. M. Walton, and C. S. Dweck, Excellence in Education: The Importance of Academic Mindsets, Excell. Educ. Importance Acad. Mindsets 42 (2013).

[47] S. Claro, D. Paunesku, and C. S. Dweck, Growth Mindset Tempers the Effects of Poverty on Academic Achievement, Proc. Natl. Acad. Sci. U. S. A. **113**, 8664 (2016).

[48] C. S. Dweck and E. L. Leggett, A Social-Cognitive Approach to Motivation and Personality, Psychol. Rev. **95**, 256 (1988).

[49] M. W. Guthrie and Z. Chen, Adding Duration-Based Quality Labels to Learning Events for Improved Description of Students' Online Learning Behavior., in *Proc. 12th Int. Conf. Educ. Data Min.*, edited by M. C. Desmarais, C. F. Lynch, A. Merceron, and R. Nkambou (International Educational Data Mining Society {(IEDMS)}, Montreal, Canada, 2019).

[50] R. Bodily, R. Nyland, and D. Wiley, The RISE Framework: Using Learning Analytics to Automatically Identify Open Educational Resources for Continuous Improvement, Int. Rev. Res. Open Distance Learn. **18**, 103 (2017).

[51] Y. J. Lee, D. J. Palazzo, R. Warnakulasooriya, and D. E. Pritchard, Measuring Student Learning with

Item Response Theory, Phys. Rev. Spec. Top. - Phys. Educ. Res. **4**, 010102 (2008).

[52] C. C. Y. Dorea, P. A. A. Resende, and C. R. Gonçalves, Comparing the Markov Order Estimators AIC, BIC and EDC, in *Trans. Eng. Technol. World Congr. Eng. Comput. Sci. 2014* (Springer Netherlands, 2015), pp. 41–54.

# VIII. Appendix: Details on Determining the Brief-Normal attempt duration cutoff

**Skewed Normal Mixture Model Fitting:** Mixture model data fitting is conducted using the four different distribution models available in the R package mixsmsn: the normal distribution; the skew-normal distribution; the skew-Student-t distribution; and the skew-contaminated normal distribution (Skew-cn). The fitting algorithm searches for the optimum number of components and fitting parameters for each distribution model according to model selection criteria EDC, which is shown to be more reliable under certain conditions [52]. The four best fit models are then compared based on four model selection criteria: AIC, BIC, EDC and ICL. The model favored by the most criteria is adopted. If more than one model is favored, then the one favored by EDC is selected.

One challenge for data fitting is that problem solving durations can be as long as several thousand seconds, whereas Brief attempts are usually under 60 seconds. Therefore, the best-fit model may be selected because of a good fit for the long duration distribution but a less accurate fit for the shorter duration, or even not able to fit the short duration at all. To prevent this, we will only use the duration distribution below a maximum duration, and increase the maximum duration from 150s to 550s at 50s intervals to examine how the maximum duration affects the estimation of the Brief-Normal cutoff distribution.

*Table 7: Best fit model and the cutoff between the shortest and the next shortest distribution, for each maximum duration cutoff analyzed.*

| MAX DURATION | MODEL | N COMP. | CUTOFF (S) |
|---|---|---|---|
| **150** | Skew.normal | 3 | 33.5 |
| **200** | Skew.cn | 3 | 46.5 |
| **250** | Skew.cn | 4 | 43.5 |
| **300** | Skew.cn | 4 | 48.5 |
| **350** | Normal | 4 | 30.5 |
| **400** | Normal | 4 | 30.5 |
| **450** | Normal | 5 | 30.5 |
| **500** | Normal | 5 | 31.5 |
| **550** | Normal | 5 | 30.5 |

The best fit model for each maximum duration, as well as the estimated cutoff between the first and second component, is listed in Table 7. For maximum durations between 200s and 300s, the multi-component skew-cn distribution is selected to be the best fit model, with the 1st cutoff estimated at around 45 s. When maximum durations are more than 350s, the multi-component normal distribution is selected as the best fit model, with 1st cutoff at around 30s. However, the normal distributions run a higher risk of over fitting, since students' problem-solving duration distribution should be skewed by nature, as there is always a minimum amount of time required to solve any problem but no a clear upper limit. Therefore, we will take 35s as our "Brief-Normal" cutoff, which is close to the average of all the cutoffs obtained for different cutoffs. As shown in Figure 6, the 35s cutoff sits right at the center of the first minimum of the distribution.

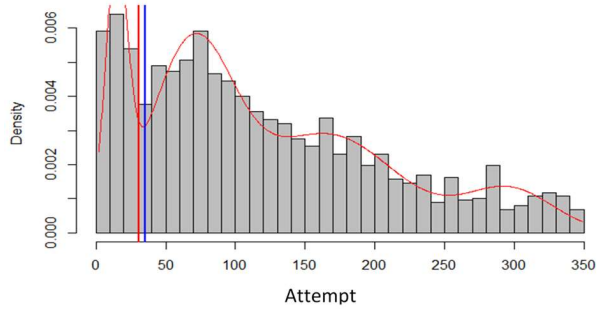

Figure 6: Example of mult-component mixture model fit of the duration distribution, with maximum duration of 350s. The red line indicates the cutoff generated by the algorithm at 30s, and the blue line indicateds the average cutoff for all durations at 35s.

**Mean log-duration of attempts before and after study:** In Table 8, we list the mean log-duration (in unit of seconds) of AC attempts both before and after studying the IC. Duration data is truncated at a maximum of 7200s. As shown in the table, modules m2 and m6 are the only two modules on which the mean log-duration reduced by more than a half from before study to after study. Therefore, the Brief-Normal cutoff on those two modules for post-study attempts are set at 17 and 24 seconds respectively for after study attempts. All other attempts used 35s as Brief-Normal cutoff.

Table 8: Mean log-duration of before and after study attempts for each module, in units of seconds.

| Attempt Type | m1 | m2 | m3 | m4 | m5 | m6 | m7 | m8 | m9 | m10 |
|---|---|---|---|---|---|---|---|---|---|---|
| Before Study | 292 | 105 | 131 | 178 | 113 | 112 | 218 | 93 | 89 | 78 |
| After Study | 276 | 35 | 143 | 111 | 118 | 53 | 300 | 211 | 108 | 82 |